\documentclass[article,prl,balancelastpage,twocolumn,showpacs]{revtex4}
\usepackage{makeidx}
\usepackage{amssymb}
\usepackage{dcolumn}
\usepackage{graphicx}
\usepackage{acronym}

\input{tcilatex}

\begin{document}

\title{ 3D transformation for point rotating reference frames and localized
neutrinos in a rotating electromagnetic field }
\author{Boris V. Gisin }
\affiliation{IPO, Ha-Tannaim St. 9, Tel-Aviv 69209, Israel. E-mail:
borisg2011@bezeqint.net}
\date{}

\begin{abstract}
The problem of Dirac's equation in a rotating electromagnetic field can be
reduced to the stationary by using a transformation for point rotating
reference frames. The general form of the non-Galilean transformation is
deduced in the paper. For the non-Galilean transformation time is different
in the rotating and resting frame. This transformation contains a constant,
with the dimension of time. The constant is assumed to be fundamental.

The transformation forms the necessary condition for the existence of
periodic, bounded and square integrable solutions in both the rotating and
resting frame. Variety of such solutions exists. Among the solutions
massless states are identified. They describe localized neutrinos with
invariable spin. The states in the resting frame are not stationary but they
have a good chance to be stable.
\end{abstract}

\pacs{02.30.Ik, 03.65.Ge, 03.65.Ta\hspace{20cm}}
\maketitle

\section{Introduction}

Solutions of the Dirac equation in a rotating electromagnetic field relate
to the non-stationary problem. The problem may turn into the stationary by a
transition to a rotating reference frame. Solutions in this frame correspond
to stationary states (Landau levels \cite{Lqm}).

The transition is realized with help of a transformation of the wave
function and coordinates. The transformation of the wave function is
uniquely determined by the electromagnetic potential, but for coordinates as
the Galilean as non-Galilean transformation may be used. Two last terms
correspond to the same and different time in the rotating and resting frame,
respectively.

Two types of rotation exist in nature. The first with one axis of rotation
pertains to mechanics. Time in such a rotating frame depends on the
frequency and distance from the axis of rotation.

The second type with the rotational axis at each point may be associated
with any rotating field. A reference frame (point rotating reference frame)
may be connected with the field. Such a frame (optical indicatrix or index
ellipsoid) is used in crystallooptics more 100 years. The concept of the
frame can be applicable not only for optics and quantum mechanics, but also
and in the general relativity \cite{S2}.

The paper reviews the second type. Coordinates of 2D transformation are
angle and time of the cylindrical coordinate system. For 3D case the
coordinate along the axis of rotation is added, while the cylindrical radius
remains unchanged. Such a frame is free of centrifugal forces.

It is well known that time in two frames, moving with different velocities,
is different. But is it valid for two frames rotating with different
frequency? From the viewpoint of modern physics the positive answer to this
question seems more preferable. Moreover the non-Galilean transformation
must contain a fundamental physical constant with the dimension of time.
This constant is lacking on the list of basis physical constants like the
speed of light, Planck's constant, electron charge, and so on. Possibly,
some problems of contemporary physics are connected with this constant. This
defines the significance of this issue.

Some theoretical and experimental aspects of non-Galilean transformation in
application to optics are presented in \cite{S2}. A simple 2D transformation
was discussed in this paper as an example. Consideration has been performed
to evaluate the time difference in the rotating frame relative to the
resting one by using existing experimental data in optics \cite{jps}.

\section{The Dirac equation}

We consider Dirac's equation\ 
\begin{equation}
i\hbar \frac{\partial }{\partial t}\Psi =c\mathbf{\alpha }(\mathbf{\mathbf{p}%
}-\frac{e}{c}\mathbf{\mathbf{A}})\Psi +\beta mc^{2}\Psi =0  \label{Dir}
\end{equation}%
in an electromagnetic field, consisting of a plane traveling circularly
polarized electromagnetic wave and a constant magnetic field, with the
potential \ 
\begin{eqnarray}
A_{1} &=&-\frac{1}{2}H_{3}y+\frac{1}{k}H\cos (\epsilon \Omega t-kz),
\label{A1} \\
A_{2} &=&\frac{1}{2}H_{3}x+\frac{1}{k}H\sin (\epsilon \Omega t-kz),
\label{A2}
\end{eqnarray}%
where $\Omega $ is the frequency, $\epsilon =+1$ and \ $\epsilon =-1$
corresponds to right and left-hand polarization, respectively, $%
k=\varepsilon \Omega /c$ is the propagation constant, $\varepsilon =+1$ and $%
\varepsilon =-1$ are used when the wave propagates along the $z$-axis and
the opposite direction, insertion of $\epsilon ,\varepsilon $ is justified
because the system is not symmetric when the signs of $\epsilon ,\varepsilon 
$ are changed independently, $\Omega $ always remains positive, $c$ is the
speed of light, $\alpha _{k},\beta $ are Dirac's matrices, $H_{3}$ is the
constant magnetic field, $H$ is the amplitude of the magnetic field of the
wave. It is well known that the amplitude of the electric and magnetic
fields is the same for the given wave. $H$ is used for comparison with the
constant magnetic field $H_{3}$.

The transformation of the wave function 
\begin{equation}
\tilde{\Psi}=\exp \frac{1}{2}\alpha _{1}\alpha _{2}(\epsilon \Omega
t-\varepsilon \frac{\Omega }{c}z)\Psi ,  \label{tpsi}
\end{equation}%
reduces Eq. (\ref{Dir}) to the stationary form. The transformation should be
accompanied by the transformation of coordinates. In a sense the problem is
similar to the problem in mechanics. For transition to the reference frame
of center mass the Lorentz transformation should be used. The reverse
transition in the laboratory frame can be performed after studies of
temporal and spatial behavior of a object.

\section{The transformation of coordinates}

The transformation of coordinates connects the cylindrical angle $\varphi ,$
coordinate along the axis of rotation $z$ and time $t$\ in the resting and
rotating frames. The cylindrical radius $r$ remains invariable. The
dependence of the angle is defined by the transformation of spinor $\tilde{%
\varphi}=\varphi -\epsilon \Omega t+\varepsilon \Omega z/c$. The dependence
of$\ \tilde{z},\tilde{t}$ on $z,t$ is taken to be a linear form of
coordinates with coefficients depending on $\Omega .$

The 3D transformation is determined by following general assumptions:

$\bullet $ the dependence of $\tilde{z},\tilde{t}$ from $z,t$ has Lorentz's
shape 
\begin{equation}
\tilde{z}=a\varphi +\frac{z+vct}{\sqrt{1-v^{2}}},\text{ }\ \tilde{t}%
=b\varphi +\frac{zv/c+t}{\sqrt{1-v^{2}}},  \label{ztL}
\end{equation}%
where $a,b,v$ are parameters depending on $\Omega .$

$\bullet $ the principle of the velocity of light constancy is kept: if the
velocity in the resting frame $V\equiv z/t=\varepsilon _{v}c$ then in the
rotating frame $\tilde{V}\equiv \tilde{z}/\tilde{t}=\varepsilon _{v}c,$
where $\varepsilon _{v}^{2}=1$. This\ condition in the general case would
give two relations between coefficients. However one relation is fulfilled
automatically, because of Lorentz's shape (\ref{ztL}), and only one remains%
\[
\text{\ }\varepsilon _{v}cb=a. 
\]

$\bullet $ the analogous principle of some frequency constancy exists: if $%
\omega \equiv \varphi /t=\epsilon _{\omega }/\tau $ then $\tilde{\omega}%
\equiv \tilde{\varphi}/\tilde{t}=\epsilon _{\omega }/\tau ,$ where $\epsilon
_{\omega }^{2}=1$ and $\tau $ is a positive constant with the dimension of
time. The condition should be valid for arbitrary $V$. From this two
relations follows%
\begin{eqnarray}
v &=&\frac{\epsilon _{\omega }\varepsilon \tau \Omega }{\sqrt{1+\tau
^{2}\Omega ^{2}}},\text{ \ }|v|<1,  \label{v} \\
b &=&\epsilon _{\omega }\tau \lbrack \epsilon _{\omega }\epsilon \tau \Omega
-(1-\sqrt{1+\tau ^{2}\Omega ^{2}})].  \label{b}
\end{eqnarray}%
The rotating frame gains the velocity $vc$ along the $z$-axis.

Finally the transformation has the form%
\begin{eqnarray}
\tilde{\varphi} &=&\varphi -\epsilon \Omega t+\varepsilon \Omega z/c, 
\nonumber \\
\text{\ }\tilde{z} &=&-\varepsilon _{v}cb\varphi +\sqrt{1+\tau ^{2}\Omega
^{2}}z+\epsilon _{\omega }\varepsilon \tau \Omega ct,  \label{trz} \\
\text{\ }\tilde{t} &=&-b\varphi +\epsilon _{\omega }\varepsilon \frac{\tau
\Omega }{c}z+\sqrt{1+\tau ^{2}\Omega ^{2}}t.  \nonumber
\end{eqnarray}

An attempt to evaluate $\tau $ is made \cite{S2} on the basis of the J. P.
Campbell, and W. H. Steier experiment \cite{jps}. Using the 2D non-Galilean
transformation $\tilde{\varphi}=\varphi -\Omega t,$ \ \ $\tilde{t}=-b\varphi
+t,$ where $b$ is a parameter depending on $\Omega $, and expanding $b$ in
power series $\ b=\tau _{0}+\tau ^{2}\Omega +\tau _{2}^{3}\Omega ^{2}+\cdots 
$, it was shown that if $\tau _{0}\neq 0$ the upper boundary for $\tau
_{0}\sim 10^{-23}\sec .$ If $\tau _{0}$ is exactly equals zero the upper
boundary for $\tau $, which can be measured optically, is $\sim
10^{-17}-10^{-18}\sec $. Even for frequencies of the order of $100GHz$ the
product $\tau \Omega <10^{-6}$. The velocity $v$ $\sim 100$ $m/\sec $ for
this value of $\tau \Omega .$

Below we operate with the upper boundary of $\tau \sim 10^{-17}$.

\section{Solutions}

We are looking for solutions of the Dirac equation. The solutions should be
periodic, bounded and square integrable perpendicularly to the z-axis. In
the rotating frame they are stationary, whereas in the resting,
non-stationary. Moreover in the paper only exact solutions are considered.

Desirable solutions in the rotating frame have following form 
\begin{equation}
\tilde{\Psi}=\exp (-\frac{i\tilde{E}}{\hbar }\tilde{t}+\frac{i\tilde{p}}{%
\hbar }\tilde{z}-in\tilde{\varphi}+D)\psi ,  \label{psin}
\end{equation}%
were $\tilde{E}$ and $\tilde{p}$ is the "energy" and "momentum" along the $z$%
-axis, $n$ is an integer, $D=-d\tilde{r}^{2}/2+d_{1}\tilde{x}+d_{2}\tilde{y}%
, $ $\tilde{x}=r\cos \tilde{\varphi},$ \ $\tilde{y}=r\sin \tilde{\varphi}$.
The wave function $\psi $ in the rotating frame\ obeys the equation%
\begin{eqnarray}
&&\{-E+i\hbar \Omega (\epsilon -\alpha _{3}\varepsilon )[\tilde{x}d_{2}-%
\tilde{y}d_{1}+\frac{1}{2}\alpha _{1}\alpha _{2}]-  \nonumber \\
&&-i\hbar c\alpha _{1}(-\tilde{x}d+d_{1})-i\hbar c\alpha _{2}(-\tilde{y}%
d+d_{2})+\alpha _{3}pc-  \nonumber \\
&&-(\alpha _{2}\tilde{x}-\alpha _{1}\text{\ }\tilde{y})e\frac{1}{2}%
H_{3}-\alpha _{1}e\frac{1}{k}H+\beta mc^{2}\}\psi =0,  \label{Eqn}
\end{eqnarray}%
where parameters $E$ and $p$ are defined as%
\begin{eqnarray}
E &=&\sqrt{1+\tau ^{2}\Omega ^{2}}\tilde{E}-\epsilon _{\omega }\varepsilon
\tau \Omega \tilde{p}c-\epsilon n\hbar \Omega ,\text{ }  \label{Et} \\
\text{\ }pc &=&\sqrt{1+\tau ^{2}\Omega ^{2}}\tilde{p}c-\epsilon _{\omega
}\varepsilon \tau \Omega \tilde{E}-\varepsilon n\hbar \Omega .  \label{pc}
\end{eqnarray}

\subsection{The condition}

A necessary condition for existence of periodic and bounded solutions is 
\begin{equation}
\tau \lbrack \epsilon _{\omega }(\sqrt{1+\tau ^{2}\Omega ^{2}}-1)+\epsilon
\tau \Omega ](\tilde{E}-\tilde{p}c\varepsilon _{v})=\hbar n,  \label{Cn}
\end{equation}%
$n$ must be an integer, because $\tilde{\varphi}$ \ is normalized so that
the variation of $\varphi \rightarrow \varphi +2\pi $ results the variation $%
\tilde{\varphi}\rightarrow \tilde{\varphi}+2\pi $ (\ref{trz}).

The reverse transition to the resting frame is realized by the
transformation coordinates and the wave function with the tilde to that
without the tilde.%
\[
-i\frac{\tilde{E}\tilde{t}}{\hbar }+i\frac{\tilde{p}\tilde{z}}{\hbar }-in%
\tilde{\varphi}=-i\frac{Et}{\hbar }+i\frac{pz}{\hbar } 
\]%
\begin{eqnarray}
\tilde{x} &=&x\cos (\epsilon \Omega t-kz)+y\sin (\epsilon \Omega t-kz),
\label{x} \\
\tilde{y} &=&y\cos (\epsilon \Omega t-kz)-x\sin (\epsilon \Omega t-kz),
\label{y}
\end{eqnarray}%
Because of the expressions for $\tilde{x},\tilde{y}$ states in resting frame
are not stationary.

Eq. (\ref{Eqn}) with definitions (\ref{Et}), (\ref{pc}) coincides with the
equation obtained by means of the Galilean transformation $\tilde{\varphi}%
=\varphi -\epsilon \Omega t+\varepsilon \Omega z/c,$ $\tilde{z}=z,$ $\tilde{t%
}=t.$ The constant\ $\tau $ and $in\tilde{\varphi}$\ disappears from Eq. (%
\ref{Eqn}). The only difference is the condition (\ref{Cn}).

Note that in a sense the optical indicatrix (index ellipsoid) of threefold
electrooptical crystal, which used in \cite{jps}, possesses some properties
of the two-component spinor. The rotation of the electric field on an angle
produces rotation of the optical indicatrix on a half of this angle.
However, there is a principal dissimilarity of the behavior the circularly
polarized light wave in the single-sideband modulator, under the action of a
rotating (modulating) electric field, and spinor in the field of rotating
electromagnetic wave. This dissimilarity is the polarization reversal at the
modulator output. In this case the constant $\tau $ is not vanished.

The even part $[b(\Omega )+b(-\Omega )]/2$ gives an asymmetry in the
frequency shift whereas the odd $[b(\Omega )-b(-\Omega )]/2$ gives the same
shift for right- and left-hand rotations of the circularly polarized light
wave. In the transformation (\ref{trz}) $b$ has as the even $-(1-\sqrt{%
1+\tau ^{2}\Omega ^{2}})$ as odd part ~$\epsilon _{\omega }\epsilon \tau
\Omega ,$ the constant $\tau _{0}=0$.

Point rotating reference frames are free from centrifugal forces. Under the
condition of (\ref{Cn}) desirable solutions are possible in both the
rotating and resting frame. This fact is a good indicator of the solution
stability.

Solutions can be classified by the form of spinor $\psi $ in (\ref{Eqn}). A
constant spinor describes the ground state. A polynomial in $\tilde{x}$, $%
\tilde{y}$ corresponds to excited states.

\subsection{Ground state}

Spinor $\psi $ of the ground state in normalized units is $\psi =N\psi
_{0}\exp \Phi ,$ 
\begin{equation}
\psi _{0}=\left( 
\begin{array}{c}
h\mathcal{E} \\ 
-\varepsilon (\mathcal{E}+\epsilon )(\mathcal{E}-\mathcal{E}_{0}) \\ 
\epsilon \varepsilon h\mathcal{E} \\ 
-\epsilon (\mathcal{E}-\epsilon )(\mathcal{E}-\mathcal{E}_{0})%
\end{array}%
\right) ,  \label{Eqg}
\end{equation}%
where the normalized constant is 
\begin{equation}
N=\frac{\sqrt{d}\exp (-d_{2}^{2}/2d)}{\sqrt{2\pi }\sqrt{h^{2}\mathcal{E}%
^{2}+(\mathcal{E}^{2}+1)(\mathcal{E}-\mathcal{E}_{0})^{2}}}.  \label{Ng}
\end{equation}%
The "normalized energy" $\mathcal{E}$ obeys the characteristic equation%
\begin{equation}
\mathcal{E}(\mathcal{E}+\Lambda )-1-\frac{\mathcal{E}h^{2}}{\mathcal{E}-%
\mathcal{E}_{0}}=0,\text{ \ }\Lambda =\frac{2\varepsilon \tilde{p}c-\hbar
\Omega }{mc^{2}},  \label{Chg}
\end{equation}%
in normalized units%
\begin{equation}
\text{ \ }\mathcal{E}\equiv \frac{E\epsilon -pc\varepsilon }{mc^{2}},\text{
\ }\mathcal{E}_{0}=\frac{2d\hbar }{\Omega m},\text{ \ }h=\frac{e}{kmc^{2}}H.
\label{Eh}
\end{equation}%
Parameters $\Phi ,d,$ $d_{1},d_{2}$ are defined as follows 
\begin{eqnarray}
\Phi  &=&-\frac{i\tilde{E}}{\hbar }\tilde{t}+\frac{i\tilde{p}}{\hbar }\tilde{%
z}-in\tilde{\varphi}-  \nonumber \\
&&-\frac{1}{2}\alpha _{1}\alpha _{2}(\epsilon \Omega t-\varepsilon \frac{%
\Omega }{c}z)-\frac{1}{2}d\tilde{r}^{2}+d_{1}\tilde{x}+d_{2}\tilde{y},
\label{Pih} \\
d &=&\pm \frac{eH_{3}}{2\hbar c}>0,\text{ \ }d_{1}=-id_{2},\text{ \ }d_{2}=%
\frac{cdh}{\Omega (\mathcal{E}-\mathcal{E}_{0})}.  \label{d012}
\end{eqnarray}

The desirable solutions are localized in the cross section with the size of
the order $l_{d}$%
\begin{equation}
l_{d}\sim \frac{1}{\sqrt{d}}=\sqrt{\frac{2\hbar c}{eH_{3}}}.  \label{ld}
\end{equation}

For definiteness we consider $d=-eH_{3}/2\hbar c.$ For the $d=+eH_{3}/2\hbar
c>0,$ the wave function is defined as$\ \psi _{+}=\varepsilon \alpha
_{1}\alpha _{3}\beta \psi _{0}$ with the simultaneous sign change of $%
\mathcal{E}_{0}$.

There exists variety of exited states, in particular, 
\begin{equation}
\psi _{e}=N_{e1}\psi _{0}(1-i\frac{d}{d_{2}}\tilde{x}-\frac{d}{d_{2}}\tilde{y%
}),  \label{Eqe2}
\end{equation}%
where $\psi _{0}$ as well as parameters $d,d_{1},d_{2}$ are the same as in (%
\ref{Eqg}) and (\ref{d012}), whereas the parameter $\Lambda $ differs $\
\Lambda =(2\varepsilon \tilde{p}c-3\hbar \Omega )/mc^{2}$.

Obviously, wave functions (\ref{Eqg}), (\ref{Eqe2}) cannot be presented as a
small and large two-component spinor. It means that the difference $%
E^{2}-m^{2}c^{2}$ cannot be small and these solutions correspond only to the
relativistic case.

Consider the condition (\ref{Cn}) in the resting frame. We obtain, with help
of (\ref{Et}), (\ref{pc}) at $\tau \Omega \ll 1$ in the first approximation 
\begin{equation}
\epsilon (E-pc\varepsilon _{v})\approx \frac{\hbar n}{\tau ^{2}\Omega }.
\label{Cn0}
\end{equation}%
For the discussed above value of $\tau \sim 10^{-17},$ $\tau \Omega \sim
10^{-6}$ the right part is of the order of $10^{3}m_{p}c^{2},$ where $m_{p}$
is of the order of\ the proton mass. For electron it corresponds huge
energy. For heavier particles this is also in line with the high energy
physics, especially, if the real value of the constant $\tau $ is less than $%
10^{-17}$.

Condition (\ref{Et}), (\ref{pc}), (\ref{Cn}) and the characteristic equation
allows to determine parameters $E$ and $\ pc$ and, consequently, the wave
function in the resting frame for $n\neq 0$. Analogously $E$ and $\ pc$ can
be calculated for $n=0$ at $\varepsilon _{v}=-\epsilon \varepsilon $..

In this paper we don't consider the case $n\neq 0$ with too cumbersome
formulas and restrict ourselves to the interesting case $n=0,$ $\varepsilon
_{v}=\epsilon \varepsilon $.

\subsection{Localized neutrinos}

Consider $\varepsilon _{v}=\epsilon \varepsilon $. The equality $\tilde{E}=%
\tilde{p}c\varepsilon _{v}$ results in the equality $\epsilon E=\varepsilon
pc$ with help of Eqs. (\ref{Et}), (\ref{pc})$.$ In this case the
characteristic equation is fulfilled only for $m=0$. The massless ground
state in these conditions is described by the wave function 
\begin{equation}
\Psi =(-\frac{d_{2}^{2}}{2d})\sqrt{\frac{d}{2\pi }}\left( 
\begin{array}{c}
0 \\ 
\varepsilon \epsilon \\ 
0 \\ 
-1%
\end{array}%
\right) \exp \Phi ,  \label{n0}
\end{equation}%
\begin{eqnarray}
\Phi &=&-i\frac{Et}{\hbar }+i\frac{pz}{\hbar }-\frac{1}{2}\alpha _{1}\alpha
_{2}(\epsilon \Omega t-\varepsilon \frac{\Omega }{c}z)-  \nonumber \\
&&-\frac{1}{2}dr^{2}+d_{1}\tilde{x}+d_{2}\tilde{y},  \label{Pih1}
\end{eqnarray}%
\begin{equation}
\text{\ }d_{2}=-\frac{\varepsilon e}{2\hbar \Omega }H.  \label{d02}
\end{equation}%
The existence of this exact solution is easy to verify by substituting (\ref%
{n0}) into the initial equation (\ref{Dir}).

Excited states also exist, in particular, the solution obtained from Eq. (%
\ref{Eqe2}).

Because solutions are not stationary it is relevant to consider the average
values of operators of energy $E_{a}$, momentum $\mathbf{p}_{a}$ and spin.
For the given case of the localization perpendicular to the $z$-axis the
average value of an operator $P$ is defined by the integral over all
cross-section%
\[
P_{a}=\int \Psi ^{\ast }P\Psi dxdy.
\]%
The integral can be calculated exactly for all average values below.

The average values of $E_{a}$ and $p_{za}$ obey the same relation as the
parameters $E$ and $p$

\begin{equation}
\epsilon E_{a}=\epsilon E-\hbar \Omega +\frac{ceH_{3}}{2\Omega }(\frac{H}{%
H_{3}})^{2}=\varepsilon p_{za}c,  \label{Ea}
\end{equation}%
The values of $p_{xa},p_{ya}$ change with $t,z$ similarly to potential $%
A_{1},A_{2}$%
\begin{eqnarray}
p_{xa} &=&\frac{\varepsilon e}{2\Omega }H\cos (\epsilon \Omega t-\varepsilon 
\frac{\Omega }{c}z),  \label{px} \\
\text{\ }p_{ya} &=&\frac{\varepsilon e}{2\Omega }H\sin (\epsilon \Omega
t-\varepsilon \frac{\Omega }{c}z).  \label{py}
\end{eqnarray}%
The average components of spin are invariable%
\begin{equation}
\text{\ }s_{3}=\frac{\hbar }{2},\text{ \ }s_{1}=s_{2}=0.  \label{sk}
\end{equation}

Other interesting massless solution follows from the previous by $E=p=0.$
The wave function, average energy and the component of momentum $p_{za}$
depends only on parameters of the electromagnetic field.

\section{Conclusion}

The 3D non-Galilean transformation for point rotating reference frames is
uniquely determined by the following general assumptions:

-Lorentz's form of the dependence of the "Cartesian coordinates" in the
rotating and resting frame;

-The principle of the velocity of light constancy;

-Similar principle of some frequency constancy.

Surprisingly the frequency, in contrast to the velocity, is not the limit
frequency in this transformation. A consequence of this transformation is
the condition of the existence of periodic, bounded and localized solutions
in the rotating and resting frame. The non-Galilean transformation contains
a constant with the dimension of time. This fundamental constant defines
processes in the very small intervals of time and length. Therefore the
transformation is of particular interest to the standards of time and length
and also for nanotechnology.

Periodic, bounded and square integrable solutions of Dirac's equation in the
rotating electromagnetic field have been considered. They describe only
relativistic fermions. These fermions are localized in the small
cross-section with the size of the order of $l_{d}$. All states that have
been found in this paper are exact solutions. Among the solutions, massless
solutions exist. These solutions are described non-stationary (but,
possibly, stable) localized neutrinos with invariable spin. Apparently, such
solutions can be used in quantum field theory and high-energy physics.

\end{document}